\documentclass[12pt]{article}
\usepackage{latexsym}
\usepackage{amssymb, amsmath, xspace, lscape,  latexsym}

\renewcommand{\le}{\leqslant}
\renewcommand{\ge}{\geqslant}

\newcommand{\pr}{\prime}
\newcommand{\non}{\nonumber}
\newcommand{\bea}{\begin{eqnarray}}
\newcommand{\eea}{\end{eqnarray}}
\def\beq#1#2\eeq{
        \begin{equation}
        \label{#1}
            #2
        \end{equation}}

\newcommand{\tp}{\textsf{p}_1}
\newcommand{\al}{\alpha}
\newcommand{\bt}{\beta}

\newcommand{\Ga}{\Gamma}
\newcommand{\re}{\textrm{e}}

\newcommand{\C}{\mathbb C}

\newcommand{\Z}{\mathbb Z}
\newcommand{\T}{\mathbb T}

\newcommand{\rex}{{\rm e}}
\renewcommand{\tilde}{\widetilde}

\newcommand{\iv}{^{-1}}
\newcommand{\iy}{\infty}
\newcommand{\bq}{\begin{eqnarray}}
\newcommand{\eq}{\end{eqnarray}}
\newcommand{\nn}{\nonumber}
\newcommand{\ba}{\begin{array}}
\newcommand{\ea}{\end{array}}
\newcommand{\ch}{\choose}

\newcommand{\PV}{\mathrm{P}_V}

\def\btheor#1\etheor{
        \begin{theor}
            #1
        \end{theor}
    }

    \def\bsled#1\esled{
        \begin{sled}
            #1
        \end{sled}   }

\newtheorem{theorem}{Theorem}

\def\hm#1{#1\nobreak\discretionary{}{\hbox{\m@th$#1$}}{}}
\def\mi#1{\discretionary{\hbox{\m@th$#1$}}{\hbox{\m@th$#1$}}{}}

\vspace{4ex}

\begin{document}
\title{\bf Painlev\'e V and time dependent Jacobi polynomials}
\author{Estelle Basor\thanks{Supported in part by NSF Grant DMS-0500892}\\
American Institute of Mathematics\\
Palo Alto, California 94306,USA\\
ebasor@aimath.org
\and
Yang Chen\\
        Department of Mathematics\\
        Imperial College London\\
        180 Queen's Gates\\
        London SW7 2BZ UK\\
        ychen@imperial.ac.uk
        \and
        Torsten Ehrhardt\\
        Department of Mathematics\\
        POSTECH\\
        Pohang 790-784, South Korea\\
        ehrhardt@postech.ac.kr}
\date{}
\maketitle
\begin{abstract}
In this paper we study the simplest deformation on a sequence of orthogonal polynomials, namely,
replacing the original (or reference) weight $w_0(x)$ (supported on $\mathbb{R}$ or
subsets of $\mathbb{R}$) by $w_0(x)\rex^{-tx}.$ It is a well-known fact that under such
a deformation the recurrence coefficients denoted as $\al_n$ and $\bt_n$ evolve in $t$
according to the Toda equations, giving rise to the time dependent orthogonal polynomials,
using Sogo's terminology.

If $w_0$ is the Gaussian density $\rex^{-x^2},\;x\in\mathbb{R},$ or the Gamma density
$x^\al \rex^{-x}$, $x\in\mathbb{R}_{+}$, $\al>-1$, then the initial value problem
of the Toda equations can be trivially solved. This is because under
elementary scaling and translation the orthogonality relations reduce to the original ones.
However, if $w_0$ is the Beta density 
$(1-x)^{\al}(1+x)^{\bt}$, $x\in[-1,1]$, $\al,\bt>-1,$
the resulting "time-dependent" Jacobi polynomials will again satisfy a linear second order ode,
but no longer in the Sturm-Louville form. This deformation
induces an irregular point at infinity in addition to three regular singular points of the
hypergeometric equation satisfied by the Jacobi polynomials.

We will show that the coefficients of this ode are intimately related to a particular
Painlev\'e V. In particular we show that $\textsf{p}_1(n,t)$, where $\textsf{p}_1(n,t)$ is
the coefficient of $z^{n-1}$ of the monic orthogonal polynomials associated with the "time-dependent" Jacobi
weight, satisfies,  up to a translation in $t,$ the Jimbo-Miwa $\sigma$-form of the same $\PV;$ while
a recurrence coefficient $\al_n(t),$
is up to a translation in $t$ and a linear fractional transformation
$\PV(\al^2/2,-\bt^2/2, 2n+1+\al+\bt,-1/2).$ These results are found from combining a pair of non-linear
difference equations and a pair of Toda equations.

This will in turn allow us to
show that a certain Fredholm determinant related to a class of Toeplitz plus Hankel
operators has a connection to a Painlev\'e equation.

The case with $\al=\bt=-1/2$ arose from a certain integrable system
and this was brought to our attention by A. P. Veselov.
\end{abstract}
\noindent
\vspace*{.5in}

\section{Introduction}
The study of Hankel determinants has seen a flurry of activity in recent years in part due to
connections with random matrix theory. In this paper we will investigate Hankel determinants for
a weight of the form
\[ (1-x)^{\al}(1+x)^{\bt}\re^{-tx} \]
on the interval $[-1,1].$ Here we take $t\in \mathbb{R}.$
We will call this a time dependent Jacobi weight.  Our ultimate goal is to produce a 
non-linear second order differential equation  that is satisfied by the logarithmic derivative of $D_n(t)$, where
$D_n(t)$ is the determinant of the Hankel matrix generated from the moments of the weight:
$$
D_n(t):=\det\left(\mu_{j+k}(t)\right)_{j,k=0}^{n-1}:=\det\left(\int_{-1}^{1}x^{j+k}
\:(1-x)^{\al}(1+x)^{\bt}\re^{-tx}dx\right)_{j,k=0}^{n-1},
$$
and we shall initially assume $\al,\:\bt >0.$

The moments $\mu_k(t)$ can be evaluated as follows:
$$
\mu_k(t)=(-1)^k\;\frac{d^k}{dt^k}\mu_0(t),\;\;k=0,1,2,...
$$
where
$$
\mu_0(t)=2^{\al+\bt+1}\Ga(\al+1)\Ga(\bt+1)\;\re^{t}\:M(\bt+1;\al+\bt+2;-2t),
$$
and $M(a;b;z)$ is the Kummer function with parameters $a$ and $b.$
Because
$$
\frac{d^k}{dz^k}M(a;b;z)=\frac{(a)_k}{(b)_k}M(a+k;b+k;z),
$$
we find,
$$
\mu_k(t)=2^{\al+\bt+1}\Ga(\al+1)\Ga(\bt+1)\:\re^{t}\:\sum_{r=0}^{k}
\:{k\ch\:r}(-2)^r\frac{(\bt+1)_r}{(\al+\bt+2)_r}\:
M(\bt+r+1;\al+\bt+r+2;2t).
$$
Our interest in evaluating $D_n(t)$ comes from the fact $D_n(t)/D_n(0)$ is the generating function of the linear
statistics
$$
\sum_{j=1}^{n}x_j,
$$
and can also be thought of as the partition function for the random matrix ensemble with eigenvalue distribution
$$
\prod_{1\leq j<k\leq n}(x_j-x_k)^2\prod_{l=1}^{n}(1-x_l)^{\al}(1+x_l)^{\bt}\:{\rm e}^{-tx_l}dx_l.
$$
The paper is
divided into the sections as follows. In the next section we reproduce known results that we call
coupled Toda equations. We include them in this paper as reference. In section 3 we consider ladder operators and
derive fundamental equations that are the basis for everything that follows. They give us
coupled difference equations and coupled Riccati equations in certain auxiliary quantities denoted as $r_n(t)$ and
$R_n(t).$

The Ricatti equations allow us to find a non-linear second order differential equation that is satisfied
by the recurrence coefficient $\al_n(t).$ A rational change of variable applied to $\al_n(t)$ is then
a solution to a Painlev\'e V in standard form and can be found in section 4.

In section 5 we identity the function which satisfies the continuous and discrete $\sigma$-form of our $\PV.$
We show that the $\sigma$-function of Jimbo, Miwa and Okamoto is given by
$$
\sigma(t)= \frac{t}{2}\textsf{p}_1(n,t/2)-\frac{n}{2}t+n(n+\bt),
$$
and since
$$
\textsf{p}_1(n,t)=\frac{d}{dt}\log D_n(t),
$$
consequently $D_n(t)$ is related in a simple way to the $\tau$-function of the $\PV.$

In section 6 we show how the Hankel
determinants can be also expressed as determinants of finite Toeplitz plus Hankel matrices.
For some special cases,
these latter determinants are known exactly and hence so are our Hankel determinants. Thus we,
in a round about way,
produce second order equations that have solutions that are logarithmic derivatives of  Fredholm determinants.
This should not come as a great surprise as this is a common occurrence
in random matrix theory for the classical ensembles.

Finally using known results for the Fredholm determinants we are able, in some special cases, to
write down asymptotic expansions for these determinants and make some predictions about higher order terms.
\setcounter{equation}{0}
\section{Preliminaries: Notations and time evolution}
The purpose of this section is to derive two coupled Toda equations that involve the recursion
coefficients of the time-dependent Jacobi polynomials. This is not a new result. The rather more general
Toda-hierarchy, can be found for example, in \cite{Moser}, \cite{Haine} and \cite{Witten}. Ours corresponds to
the first of the hierarchy. See \cite{Haine+S} for a discussion of this in relation to Sato's theory.
See also \cite{A+V} for the ``multi-time'' approach to matrix models.

We include the necessary computations here for completeness
sake and to set the notations to be used throughout this paper.

To begin we consider general orthogonal polynomials $P_{i}(x)$ with respect to the weight
$w_0(x) \rex^{-tx}$ on $[-1,1].$ The weight $w_{0}$ will be known as the ``reference'' weight.
The orthogonality condition is
\bea
\int_{-1}^{1}P_i(x)P_j(x)w_0(x)\re^{-tx}dx=h_i(t)\delta_{i,j},
\eea
and the $t$ dependence through $\re^{-tx},$ induces $t$ dependence
on the coefficients. We normalize our monic polynomials as
\bea
P_n(z)=z^n+\textsf{p}_{1}(n,t)z^{n-1}+...+P_n(0),
\eea
although sometime we do not display the $t$ dependence of coefficients of $z^{n-1}$.

An immediate consequence of the orthogonality condition is the three terms recurrence relation
\bea
zP_{n}(z)=P_{n+1}(z)+\al_nP_n(z)+\bt_{n}P_{n-1}(z)
\eea
with the initial conditions
\bea
P_0(z)=1,\;\;\bt_0P_{-1}(z)=0.
\eea
An easy consequence of the recurrence relation is
\bea
\al_n(t)=\textsf{p}_1(n,t)-\textsf{p}_1(n+1,t),
\eea
and a telescopic sum of the above equation (bearing in mind that $\textsf{p}_1(0,t)=0)$ leaves
\bea
-\sum_{j=0}^{n-1}\al_j(t)=\textsf{p}_1(n,t).
\eea
First let us discuss the derivatives of $\al_n$ and $\bt_n$ with
 respect to $t,$ as
this yields the simplest equations, where we keep $w_0$ quite general, as long as
the moments
\bea\label{f.mu-i}
\mu_i(t):=\int_{-1}^{1}x^{i}w_0(x)\re^{-tx}dx,\quad i=0,1,...
\eea
of all orders exist. Taking a derivative of $h_n$ with respect to $t$
\bea
h_n'(t)=-\int_{-1}^{1}w_0(x)\re^{-tx}xP_n^2(x)dx=-\al_n h_n,
\eea
i.e.,
\bea
(\log h_n)'=-\al_n,
\eea
and since $\bt_n=h_n/h_{n-1},$ we have
the first Toda equation,
\bea
\bt_n'=(\al_{n-1}-\al_n)\bt_n.
\eea
We define $D_{n}(t)$ to be the Hankel determinant
\bea\label{f.Dn-t}
D_n(t)= \det (\mu_{i+j}(t))_{i,j=0}^{n-1} .
\eea
It is well-known that $D_{n}(t)= \prod_{ i= 0}^{n-1} h_{i}(t) .$
This yields in view of (2.9) that
\bea
\frac{d}{dt}\log D_n(t)=-\sum_{j=0}^{n-1}\al_j(t)=\textsf{p}_1(n , t).\non
\eea
Also,
\bea
0&=&\frac{d}{dt}\int_{-1}^{1}P_nP_{n-1}w_0\re^{-tx}dx\nonumber\\
&=&-\int_{-1}^{1}xP_nP_{n-1}w_0\re^{-tx}dx+h_{n-1}\;\frac{d}{dt}\textsf{p}_1(n,t)\nonumber\\
&=&-h_n+h_{n-1}\;\frac{d}{dt}\textsf{p}_1(n,t),\non
\eea
and therefore
\bea
\frac{d}{dt}\textsf{p}_1(n,t)=\bt_n(t).
\eea
But since $\al_n=\textsf{p}_1(n) -\textsf{p}_1(n+1),$ we have the second Toda equation,
\bea
\al_n'=\bt_n-\bt_{n+1}.
\eea
To summarize we have the following theorem.
\begin{theorem}
The recursion coefficients $\al_{n}(t)$ and $\bt_{n}(t)$ satisfy the coupled
Toda equations,
\bea
\bt_n'=(\al_{n-1}-\al_n)\bt_n,
\eea
\bea
\al_n'=\bt_n-\bt_{n+1}.
\eea
\end{theorem}
It is also worth pointing out that in view of (2.11) we have the obvious Toda molecule equation \cite{Sogo},
\bea
\frac{d^2}{dt^2}\log D_n(t)=\frac{d}{dt}\textsf{p}_1(n,t)=\bt_n(t)
=\frac{D_{n+1}(t)D_{n-1}(t)}{D_n^2(t)}.\non
\eea
\setcounter{equation}{0}
\section{Ladder operators, compatibility conditions, and difference equations.}

In this section we give an account for a recursive algorithm for the
determination of the recurrence coefficients $\al_n,\;\;\bt_n$ based a pair of ladder operators
and the associated supplementary conditions. Such operators have been derived by
various authors over many years. Here we provide a brief guide to the relevant literature,
\cite{Bauldry}, \cite{Belmehdi}, \cite{Bonan+Nevai}, \cite{Bonan+Calrk1}, \cite{Bonan+Clark2},
\cite{chen1}, \cite{chen2}, \cite{F+I+K1}, \cite{F+I+K2} and \cite{Mag1}. In fact Magnus
in \cite{Mag1} traced this back to Laguerre. We find the form of the ladder operators set out
below convenient to use.

For a sufficiently well-behaved  weight (see \cite{chen2} for a precise statement) of the form
$w(x)={\rm e}^{-\textsf{v}(x)}$ the lowering and raising operators are
\bea
P_n^{\pr}(z)&=&-B_n(z)P_n(z)+\bt_nA_n(z)P_{n-1}(z),\\
P_{n-1}^{\pr}(z)&=&[B_n(z)+\textsf{v}^{\pr}(z)]P_{n-1}(z)-A_{n-1}(z)P_n(z),
\eea
where
\bea
A_n(z)&:=&\frac{1}{h_n}\int_{-1}^{1}
\frac{\textsf{v}'(z)-\textsf{v}'(y)}{z-y}
P_n^2(y)w(y)dy,\\
B_n(z)&:=&
\frac{1}{h_{n-1}}\int_{-1}^{1}\frac{\textsf{v}'(z)-\textsf{v}'(y)}{z-y}
P_n(y)P_{n-1}(y)w(y)dy.\\
\eea
Here we have assumed that $w(\pm 1)=0.$ Additional terms would have to be included in the definitions of
$A_n(z)$ and $B_n(z)$ if $w(\pm 1)\neq 0.$ See \cite{chen1} and \cite{chen2}.

A direct calculation produces two fundamental supplementary (compatibility) conditions valid
for all $z;$
$$
B_{n+1}(z)+B_n(z)=(z-\al_n)A_n(z)-\textsf{v}'(z)\eqno(S_1)
$$
$$
1+(z-\al_n)(B_{n+1}(z)-B_n(z))=\bt_{n+1}A_{n+1}(z)-\bt_nA_{n-1}(z).
\eqno(S_2)
$$
We note here that $(S_1)$ and $(S_2)$ have been applied to random matrix theory in \cite{T+W}.
It turns out that there is an equation which gives better insight into the
$\al_n$ and $\bt_n$ if $(S_1)$ and $(S_2)$ are suitably combined. See \cite{Chen+Its}.

Multiplying $(S_2)$ by $A_n(z)$ we see that the r.h.s. of the resulting equation is
a first order difference, while the l.h.s., with $(z-\al_n)$ replaced by
$B_{n+1}(z)+B_{n}(z)+\textsf{v}^{\pr}(z),$ is a first order difference plus $A_n(z)$.
Taking a telescope sum together with the appropriate ``initial condition'',
$$B_0(z)=A_{-1}(z)=0,$$ produces,
$$
B^2_n(z)+\textsf{v}'(z)B_n(z)+\sum_{j=0}^{n-1}A_j(z)=\bt_nA_n(z)A_{n-1}(z).
\eqno(S'_2).
$$
This last equation will be highly useful in what follows.
The equations $(S_1),$ $(S_2)$ and $(S_2')$ were also stated to \cite{Mag1}. See also \cite{for}.

If $w_0$ is modified by the multiplication of  ``singular'' factors such as $|x-t|^{a}$ or $a+bH(x-t),$ where $H$ is the unit
step function, then the ladder operator relations, $(S_1),$ $(S_2)$ and $(S_2')$ remain valid with appropriate adjustments.
See \cite{chen+p}, \cite{chen+f}, and \cite{Basor+Chen}.

 Let $\Psi(z)=P_n(z)$. Eliminating $P_{n-1}(z)$ from the raising and lowering operators gives
\bea \label{2order.ode}
\Psi''(z)-\left(\textsf{v}^{\pr}(z)+\frac{A_n'(z)}{A_n(z)}\right)\Psi'(z)
+\left(B_n'(z)-B_n(z)\frac{A_n'(z)}{A_n(z)}+\sum_{j=0}^{n-1}A_j(z)\right)\Psi(z)=0,
\eea where we have used $(S_2')$ to simplify the coefficient of
$\Psi$ in (\ref{2order.ode}).

For the problem at hand, $w(x)$ is the "time-dependent" Jacobi weight, and now we must suppose that
$\al > 0$ and $\bt >0$ so that our weight is suitably well-behaved. Then
\bea
w(x)&:=&(1-x)^{\al}(1+x)^{\bt}\re^{-tx},\quad x\in[-1,1],\\
\textsf{v}(z)&:=&-\al\log(1-z)-\bt\log(1+z)+tz,\non\\
\textsf{v}'(z)&=&-\frac{\al}{z-1}-\frac{\bt}{z+1}+t,\non\\
\frac{\textsf{v}'(z)-\textsf{v}'(y)}{z-y}&=&\frac{\al}{(y-1)(z-1)}+\frac{\bt}{(y+1)(z+1)}.
\eea
Substituting these into the definitions of $A_n(z)$ and $B_n(z)$ and integrating by parts, produces,
\bea
A_n(z)&=&-\frac{R_n(t)}{z-1}+\frac{t+R_n(t)}{z+1},\non\\
B_n(z)&=&-\frac{r_n(t)}{z-1}+\frac{r_n(t)-n}{z+1},\non
\eea
where
\bea
R_n(t)&:=&\frac{\al}{h_n}\int_{-1}^{1}\frac{P_n^2(y)}{1-y}\;(1-y)^{\al}(1+y)^{\bt}\re^{-ty}dy,
\non\\
r_n(t)&:=&\frac{\al}{h_{n-1}}\int_{-1}^{1}\frac{P_n(y)P_{n-1}(y)}{1-y}\;
(1-y)^{\al}(1+y)^{\bt}\re^{-ty}dy.\non
\eea
Substituting the expressions for $A_n(z)$ and $B_n(z)$ into $(S_1)$ and $(S_2')$, which are
identities in $z$, and equating
the residues of the poles at $z=\pm 1,$ we find four distinct difference equations and one which
importantly performs the summation $\sum_jR_j:$
\bea \label{ex1}
-(r_{n+1}+r_n)&=&\al-R_n(1-\al_n)
\eea
\bea \label{ex2}
r_{n+1}+r_n&=&2n+1+\bt-(R_n+t)(1+\al_n)
\eea
\bea \label{ex3}
r_n^2+\al r_n&=&\bt_nR_nR_{n-1} 
\eea
\bea \label{ex4}
(r_n-n)^2-\bt(r_n-n)&=&\bt_n(R_n+t)(R_{n-1}+t)
\eea
\bea \label{ex5}
\left(\frac{\bt-\al}{2}\right)r_n+\frac{\al n}{2}&-&t\:r_n-r_n(r_n-n)-\sum_{j=0}^{n-1}R_j\\ \non
&=&-\frac{\bt_n}{2}[R_n(R_{n-1}+t)+(R_n+t)R_{n-1}].
\eea
We now manipulate the equations (\ref{ex1})--(\ref{ex5}) with the aim of expressing
 the recurrence coefficients $\al_n,\;\bt_n$ in terms of $r_n,\;R_n$, and of course
$n,\;t.$

\noindent
Adding (\ref{ex1}) and (\ref{ex2}) yields,
\bea
2R_n=2n+\al+\bt+1-t-t\al_n,
\eea
thus, $\al_n$ is ``easily'' expressed in terms of $R_n.$

Subtracting (\ref{ex1}) from (\ref{ex2}) gives,
\bea
r_{n+1}+r_n=n+\frac{\bt-\al+1-t}{2}-\left(\frac{t}{2}+R_n\right)\al_n.
\eea
Eliminating $\bt_nR_nR_{n-1}$ from (\ref{ex3}) and (\ref{ex4}) we find,
\bea
n(n+\bt)-(2n+\al+\bt)r_n=\bt_n[t^2+t(R_{n-1}+R_n)].
\eea
Now with the aid of (3.11), replacing $\bt_nR_{n-1}$ by $(r_n^2+\al r_n)/R_n,$
in (3.16) we find,
\bea
t(t+R_n)\bt_n=n(n+\bt)-(2n+\al+\bt)r_n-\frac{t}{R_n}(r_n^2+\al r_n),
\eea
and this expresses $\bt_n$ in terms of $R_n,\;r_n,\;n,\;t$. It is important to note the absence of
$R_{n\pm 1},$ and $r_{n\pm 1}.$
The reader will notice that the above manipulations prove that we have expressed
 $\al_{n}$ and $\bt_{n}$ the recurrence coefficients in terms of auxiliary quantities $r_{n}$ and $R_{n}$.

 This is summarized in:
\begin{theorem} With $r_{n}$, $R_{n}$, $\al_{n}$ and $\bt_{n}$ as defined above and with $\al, \bt > 0 $
\bea \label{th2.1}
t\al_n=2n+1+\al+\bt-t-2R_n,
\eea
\bea \label{th2.2}
t(t+R_n)\bt_n=n(n+\bt)-(2n+\al+\bt)r_n-\frac{t}{R_n}(r_n^2+\al r_n),
\eea
\end{theorem}
In what follows we will find two Riccati equations, one in $r_{n}$ with coefficients involving $R_{n}$
and another  with the roles of $r_{n}$ and $R_{n}$ reversed.
We will first show that
\bea
\frac{d}{dt}\log D_n(t)=n-2r_n-\bt_n t
=-\sum_{j=0}^{n-1}\al_j(t)=\textsf{p}_1(n,t).
\eea
The equation (3.20) can be derived as follows:
Replace $n$ by $j$ in (\ref{th2.1}) and sum over $j$ from 0 to $n-1$ to obtain
$$
\sum_{j=0}^{n-1}R_j=\frac{n(n-1)}{2}+\frac{n(\al+\bt+1-t)}{2}
-\frac{t}{2}\sum_{j=0}^{n-1}\al_j.
$$
 Now we can obtain from (\ref{ex5}) another expression for
 $\sum_{j}R_j,$
\bea
\sum_{j=0}^{n-1}R_j&=&\left(\frac{\bt-\al}{2}\right)r_n+\frac{\al n}{2}-tr_n-r_n^2+nr_n
+\frac{t\bt_n}{2}(R_n+R_{n-1})+r_n^2+\al r_n\non\\
&=&\left(\frac{\al+\bt}{2}+n-t\right)r_n+\frac{\al n}{2}+\frac{\bt_nt}{2}(R_n+R_{n-1})
\non\\
&=&\frac{n(n+\al+\bt)}{2}-t\:r_n-\frac{\bt_nt^2}{2},
\eea
where we have eliminated $t\bt_n(R_{n}+R_{n-1})$ using (3.16).
These last two equations yield
$$
\textsf{p}_1(n,t)=-\sum_{j=0}^{n-1}\al_j=n-2r_n-t\bt_n.
$$
>From this we also deduce (see (2.5)),
\bea
\al_n=2(r_{n+1}-r_n)+t(\bt_{n+1}-\bt_n)-1.
\eea
Hence, in view of (2.12),
\bea
\al_n=2(r_{n+1}-r_n)-t\frac{d\al_n}{dt}-1,\non\\
\eea 
i.e.,
\bea t\frac{d\al_n}{dt}+\al_n+1=
2\left(n+\frac{\bt-\al+1-t}{2}-\left(\frac{t}{2}+R_n\right)\al_n-2r_n\right),\non
\eea where we have replaced $r_{n+1}$ by $r_n$ plus an additional
term with (3.15). Finally, \bea
t\frac{d\al_n}{dt}+\al_n=2n+\bt-\al-t-(t+2R_n)\al_n-4r_n,\non \eea
or replacing $\al_n$ in favor of $R_n$ from (3.18),
\bea \label{ricotti.1}
t\frac{dR_n}{dt}=\al t+(2n+1+\al+\bt-2t)R_n-2R^2_n+2tr_n.
\eea
This is a Riccati equation in  $R_n$, but with $r_n$
appearing linearly.

\vspace*{.25in}
\noindent
Now since
$$
-\sum_{j=0}^{n-1}\al_j(t)=\textsf{p}_1(n,t)\;\; {\rm and\;}\quad \textsf{p}_1'(n,t)=\bt_n(t),
$$
see (2.6) and (2.11), we find, upon taking a derivative of (3.20) with respect to $t,$
\bea
-2\frac{dr_n}{dt}-\bt_n-t\frac{d\bt_n}{dt}=\bt_n,\non
\eea
or
\bea
-\frac{dr_n}{dt}=\bt_n+\frac{t}{2}(\al_{n-1}-\al_n)\bt_n,
\eea
where we have replaced $\bt_n'(t)$ by $(\al_{n-1}-\al_n)\bt_n$ with the aid
of the first Toda equation, (2.13).
A simple computation with (3.14) gives,
\bea
\frac{t}{2}(\al_{n-1}-\al_n)=-1+R_n-R_{n-1},\non
\eea
hence
\bea \label{ricotti.2}
-\frac{dr_n}{dt}&=&\bt_n(R_n-R_{n-1})\non\\
&=&\bt_nR_n-\frac{r_n^2+\al r_n}{R_n}\\
&=&\frac{R_n}{t(t+R_n)}
\left[n(n+\bt)-(2n+\al+\bt)r_n-\frac{t}{R_n}\left(r_n^2+\al r_n\right)\right]
-\frac{r_n^2+\al r_n}{R_n},\non
\eea
which is a Riccati equation in $r_n.$
We summarize in the following theorem:

\begin{theorem} The quantities $r_{n}$ and $R_{n}$ satisfy the coupled Riccati equations:
\bea
t\frac{dR_n}{dt}&=&\al t+(2n+1+\al+\bt-2t)R_n-2R^2_n+2r_nt,\\
-\frac{dr_n}{dt}&=&\frac{R_n}{t(t+R_n)}
\left[n(n+\bt)-(2n+\al+\bt)r_n-\frac{t}{R_n}\left(r_n^2+\al r_n\right)\right]
-\frac{r_n^2+\al r_n}{R_n}\quad
\eea
\end{theorem}
\vskip 1cm
We end this section by pointing out that the above equations not only produce
differential equations in our various unknown quantities, but also a pair of coupled non-linear
first order difference equations in $R_n$ and $r_{n}.$ If we substitute $\bt_{n}$ into (3.11) we obtain
the following result.

\begin{theorem} 
The quantities $r_{n}$ and $R_{n}$ satisfy the coupled difference equations
\bea
2t(r_{n+1}+r_n)&=& 4R_{n}^{2} + 2R_{n}(2t -2n -1 -\al -\bt) -2\al t\\
n(n+\bt)-(2n+\al+\bt)r_n & = &(r_n^2+\al r_n)\left(\frac{t^2}{R_nR_{n-1}}
+\frac{t}{R_n}+\frac{t}{R_{n-1}}\right)
\eea
together with the ``initial''  conditions
\bea
r_0&=&0,\\
R_0&=&\frac{\al+\bt+1}{2}\frac{M(1+\bt;\al+\bt+1;-2t)}{M(1+\bt;\al+\bt+2;-2t)}.
\eea
\end{theorem}
The initial condition for $R_{0}$ can be found by direct integration.
Observe that the representation of $R_0$ in terms of a ratio of the Kummer functions
allows for the analytic continuation of $\al,\:\bt$ down to $\al=\bt=-1/2,$ due to the
relation,
$$
\lim_{b\to 0}b\:M(a;b;z)=a\:zM(a+1,2,z).
$$
Hence
$$
\lim_{\al\to -1/2,\bt\to -1/2}R_0(t)=\frac{t}{2}\left(\frac{I_1(2t)}{I_0(2t)}-1\right).
$$
Indeed, by formally continuing $\bt$ so that $\bt+1=-k,\;k=0,1,2...,$ we find
$$
R_0(t)=\frac{\al}{2}\frac{L_k^{(\al-1-k)}(-2t)}{L_k^{(\al-k)}(-2t)},
$$
expressed as ratio of Laguerre polynomials of degree $k.$ It is clear that iterating (3.30)
and (3.31) with the above $R_0$ and $r_0=0$ will generate rational solutions (in the variable $t$)
of our $\PV$ derived in section 4. It is interesting to note that $R_0$ is also a rational function of
$\al$ and $t,$ therefore for the values of the parameters stated above our $\PV$ is a rational function in
$\al$ and $t.$

Also note that the above equations define the quantities $r_{n}$ and $R_{n}$ for all $\al > -1 $
and $\bt > -1$.
To verify our answers we return to the pure Jacobi case and let  $t=0,$ then (3.14) gives,
$$
R_n=n+\frac{\al+\bt+1}{2},
$$
and is consistent with (3.21) at $t=0.$ Now equating (3.11) and (3.12) at $t=0,$
gives,
$$
r_n=\frac{n(n+\bt)}{\al+\bt+2n},
$$
and
$$
\bt_n=\frac{r_n^2+\al r_n}{R_{n}R_{n-1}}=
\frac{4n(n+\al)(n+\bt)(n+\al+\bt)}{(2n+\al+\bt+1)(2n+\al+\bt-1)(2n+\al+\bt)^2}.
$$
With the $R_n$ given above, we find $\al_n$ from (3.14) at $t=0,$
$$
\al_n=\frac{\bt^2-\al^2}{(2n+\al+\bt)(2n+\al+\bt+2)},
$$
which are in agreement with those of \cite{chen2}.

\section{Identification of $\boldsymbol{\PV}$}

The idea is to eliminate $r_n(t)$ from our coupled Ricatti equations to produce
a second order ode in $R_n(t).$ This is straightforward and messy and we omit the details.
A further change of variable,
$$
R_n(t)=-\frac{t}{1-y(t)},
$$
leaves, after some simplification,
$$
y''=\frac{3y-1}{2y(y-1)}\:(y')^2-\frac{y'}{t}+{\rm Rat}(y,t),
$$
where the last term  is a particular rational function of two variables defined
as follows:
\bea
{\rm Rat}(y,t)
=-\frac{2y(y+1)}{y-1}+\frac{2(2n+1+\al+\bt)y}{t}+\frac{(y-1)^2}{t^2}\left[\frac{\al^2}{2}y-\frac{\bt^2}{2y}
\right].
\nonumber
\eea
Therefore,
$$y(t):=1+\frac{t}{R_n(t)},$$
satisfies,
$$
y''=\frac{3y-1}{2y(y-1)}(y')^2-\frac{y'}{t}+2(2n+1+\al+\bt)\frac{y}{t}
-2\frac{y(y+1)}{y-1}+\frac{(y-1)^2}{t^2}\left[\frac{\al^2}{2}y-\frac{\bt^2/2}{y}\right],
$$
which is almost a $\PV.$ To fit the above  to into a $\PV,$
we make the replacement $t\rightarrow t/2$ followed by
$$
y(t/2)=Y(t),
$$
and find
\bea
Y''&=&\frac{3Y-1}{2Y(Y-1)}(Y')^2-\frac{Y'}{t}
+\frac{(Y-1)^2}{t^2}\left[\frac{\al^2}{2}Y-\frac{\bt^2/2}{Y}\right]\non\\
&+&(2n+1+\al+\bt)\:\frac{Y}{t}
-\frac{1}{2}\frac{Y(Y+1)}{Y-1},
\eea
which is
$$\PV(\al^2/2,-\bt^2/2,2n+1+\al+\bt,d=-1/2).$$
The initial conditions are
\bea
Y(0)=1,\quad Y'(0)=\frac{1}{2n+\al+\bt+1}.\non
\eea
It is well known that there is a Hamiltonian associated with $\PV.$ To identify it, we substitute
\bea
R_n(t)&:=&-tq\\
r_n(t)&:=&-pq(q-1)+\rho\;q,
\eea
where $p=p(t),\;q=q(t)$ into (3.20), and choose $\rho$ so that the resulting expression is a
polynomial in $p$ and $q.$ There are two possible $\rho:$ $\rho=n$ and $\rho=n+\bt.$
\vskip .3cm
\noindent
{\bf Case~I.~$\rho=n$}
\bea
t\textsf{p}_1(n,t)+n(n+\al+\bt)-nt
=p(p+2t)q(q-1)-2ntq+\bt\:pq+\al\:p(q-1)
\eea
\noindent
{\bf Case~II.~$\rho=n+\bt$}
\bea
\lefteqn{t\textsf{p}_1(n,t)+n(n+\al+\bt)-\al\bt-nt}\hspace*{14ex}
\non\\
&=&p(p+2t)q(q-1)+2(\bt-n)qt+\bt pq+\al p(q-1).\qquad
\eea
Replacing $t$ by $t/2$ we see that the l.h.s. of (4.36) and (4.37) are the two Hamiltonian $t\textsf{H}_1$
and $t\textsf{H}_{2}$ for our $\PV.$
The Hamiltonian as presented in Okamoto \cite{O} (see also \cite{sakai})  is
$$
t\textsf{H}=p(p+t)q(q-1)+\al_2qt-\al_3pq-\al_1p(q-1),
$$
where
$$
a=\frac{\al_1^2}{2},\quad b=-\frac{\al_3^2}{2},\quad c=\al_0-\al_2,\quad d=-\frac{1}{2},
\quad \al_0=1-\al_1-\al_2-\al_3.
$$
Comparing with our $t\textsf{H}_{1},$ while keeping in mind that the $t$ is in fact $t/2,$
we find
\bea
\al_2&=&-n,\quad\al_3=-\bt,\quad\al_1=-\al,\quad\al_0=n+1+\al+\bt,\nonumber\\
a&=&\frac{\al^2}{2},\quad b=-\frac{\bt^2}{2},\quad c=2n+1+\al+\bt.
\eea
Comparing with our $t\textsf{H}_{2},$ we find
\bea
\al_2&=&\bt-n,\quad\al_3=-\bt,\quad\al_1=-\al,\quad\al_0=n+1+\al\nonumber\\
a&=&\frac{\al^2}{2},\quad b=-\frac{\bt^2}{2},\quad c=2n+1+\al+\bt.
\eea
Hence both $\textsf{H}_1$ and $\textsf{H}_{2}$ generates our $\PV,$
where
$$
Y(t)=1-\frac{1}{q(t/2)}.
$$
\setcounter{equation}{0}
\section{The continuous and discrete $\boldsymbol{\sigma}$-form of $\boldsymbol{\PV}$.}
Recall from the section 2, that
$$
\frac{d}{dt}\log D_n(t)=\textsf{p}_1(n,t),
$$
and
\bea
\textsf{p}_1'(n,t)=\bt_n(t).
\eea
Now we come the continuous $\sigma$-form of $\PV$ satisfied by
$\textsf{p}_1(n,t),$ with $n$ fixed and $t$ being the variable.
The idea is to express $\bt_n,\;r_n,\;r_n'$ in terms of $\tp(n,t)$ and its derivative
with respect to $t.$ Let us begin with (3.20),
\bea
\textsf{p}_1(n,t)
&=&n-2r_n-t\bt_n\nonumber\\
&=&n-2r_n-t\tp'(n,t).
\eea
>From the last equality of (5.2) we have
\bea
r_n(t)&=&\frac{1}{2}\left[n-\frac{d}{dt}(t\textsf{p}_1(n,t))\right].
\eea
Under some minor rearrangements, equations (\ref{th2.2}) and (\ref{ricotti.2}) become
\bea
t\bt_nR_n+\frac{t}{R_n}(r_n^2+\al r_n)&=&n(n+\bt)-(2n+\al+\bt)r_n-t^2\bt_n\\
-t\bt_nR_n+\frac{t}{R_n}(r_n^2+\al r_n)&=&tr_n',
\eea
respectively, which is a system of linear equations in $1/R_n$ and $R_n.$
Solving for $1/R_n$ and $R_n$ we find,
\bea
\frac{2t}{R_n}(r_n^2+\al r_n)&=&tr_n'+n(n+\bt)-(2n+\al+\bt)r_n -t^2\bt_n\nonumber\\
2t\bt_nR_n&=&-tr_n'+n(n+\bt)-(2n+\al+\bt)r_n-t^2\bt_n.
\eea
Taking the product of the above we arrive at an identity free of $R_n:$
\bea
4t^2\bt_n(r_n^2+\al r_n)=[n(n+\bt)-(2n+\al+\bt)r_n-t^2\bt_n]^2-(tr'_n)^2.
\eea

To identify the $\sigma$-function of Jimbo and Miwa \cite{J+M}, we replace $t$ by $t/2$ so that
$$
R_n(t/2)=-\frac{t}{2(1-Y(t))},
$$
and substitute the above in (3.27) in the variable $t/2.$ After a little simplification we find
$$
t\frac{dY}{dt}=tY-2r_n(t/2)(1-Y)^2-(Y-1)(\al Y+ 2n+\bt).
$$
Comparing this with the first equation of (C.40) of \cite{J+M}, we have
\bea
z(t)&=&-r_n(t/2),\nonumber\\
\frac{\theta_0-\theta_1+\theta_{\infty}}{2}&=&\al,\nonumber\\
\frac{3\theta_0+\theta_1+\theta_{\infty}}{2}&=&-2n-\bt,\nonumber
\eea
and consequently
$$
1-\theta_0-\theta_1=2n+1+\al+\bt=c,
$$
consistent with the parameter $c$ of our $\PV.$ Furthermore comparing (4.33) with (C.41) of \cite{J+M},
we find a possible identification;
\bea
\al&=&\frac{\theta_0-\theta_1+\theta_{\infty}}{2}\nonumber\\
\bt&=&\frac{\theta_0-\theta_1-\theta_{\infty}}{2},\nonumber
\eea
and consequently,
$$
\theta_0=-n,\quad\theta_1=-n-\al-\bt,\quad\theta_{\infty}=\al-\bt.
$$
But since
$$
\frac{d}{dt}\sigma(t)=z(t)=-r_n(t/2),
$$
and bearing in mind (5.3) we have, upon integration and fixing a constant,
\bea
\sigma(t)=\frac{1}{2}t\textsf{p}_1(n,t/2)-\frac{nt}{2}+n(n+\bt).
\eea
The $\sigma$-form of our $\PV$ is essentially a second order non-linear ode satisfied by
$\textsf{p}_1(n,t),$ and reads
\bea
(t\sigma'')^2=[\sigma-t\sigma'+(2n+\al+\bt)\sigma']^2+
4[\sigma-n(n+\bt)-t\sigma'][(\sigma')^2-\al\sigma'],
\eea
with the initial conditions
\bea
\sigma(0)=n(n+\bt),\quad \sigma'(0)=-r_n(0)=-\frac{n(n+\bt)}{\al+\bt+2n}.\non
\eea
After some calculations we find that (5.9) in fact the Jimbo-Miwa $\sigma$-form (C.45) with
\bea
\nu_0=0,\quad\nu_1=-\al,\quad\nu_2=n,\quad\nu_3=n+\bt.
\eea
To obtain (5.9) we first replace $t$ by $t/2$ in (5.7), and substitute
\bea
r_n(t/2)&=&-\sigma'(t)\nonumber\\
\frac{d}{dt}r_n(t/2)&=&-\sigma''(t)\nonumber\\
\bt_n(t/2)&=&2\frac{d}{dt}\textsf{p}_1(n,t/2)=4\frac{d}{dt}\frac{\left[\sigma(t)+nt/2-n(n+\bt)\right]}{t}\nonumber
\eea
into (5.7) at $t/2.$ Furthermore, since $\textsf{p}_1(n,t)=(\log D_n(t))',$ we have
$$
D_n(t)=D_n(0)\exp\left[\int_{0}^{t}\frac{\sigma(2s)-n(n+\bt)+ns}{s}ds\right],
$$
where $D_n(0)$ given by (1.6) of \cite{Basor+Chen1}.

We expect there exists a discrete analog of the continuous $\sigma$-form, namely, a difference equation
in the variable $n,$ satisfied by $\tp(n,t)$ with $t$ fixed. To simplify notations, we do not
display the $t$ dependence. The idea is similar to the continuous case namely, we express $\bt_n,$ $r_n$ and
$R_n$ in terms of $\tp(n)$ and $\tp(n\pm 1),$ and substitute these into (3.11) that is,
$$
r_n^2+\al\:r_n=\bt_n R_n\:R_{n-1}.
$$
To begin with, we note that (3.20) is linear in $\bt_n$ and $r_n,$ which
we re-write as;
\bea
t\bt_n+2r_n=n-\tp(n).
\eea
We now find a another linear equation in $\bt_n$ and $r_n.$ First note that
$\al_n=\tp(n)-\tp(n+1),$ and from (3.14) we have,
\bea
2R_n+t=2n+1+\al+\bt+t[\tp(n+1)-\tp(n)].
\eea
The sum (5.12) at ``$n$'' and the same but at ``$n-1$'', leaves,
$$
R_n+R_{n-1}+t=(2n +\al+\bt)+(t/2)[\tp(n+1)-\tp(n-1)].
$$
Substituting the above into (3.16) results the other linear equation mentioned above;
\bea
t\bt_n\{2n+\al+\bt+(t/2)[\tp(n+1)-\tp(n-1)]\}+(2n+\al+\bt)r_n=n(n+\bt).
\eea
Solving for $t\bt_n$ and $2r_n$ from the linear system (5.12) and (5.13), we find,
\bea
2r_n&=&\frac{(t/2)[n-\tp(n)][\tp(n+1)-\tp(n-1)]-n(n+\bt)}{n+(\al+\bt)/2+(t/2)[\tp(n+1)-\tp(n-1)]}\\
t\bt_n&=&\frac{[n-\tp(n)][n+(\al+\bt)/2]+n(n+\al+\bt)}{n+(\al+\bt)/2+(t/2)[\tp(n+1)-\tp(n-1)]},
\eea
and the discrete $\sigma$-form results from substituting (5.12), (5.14) and (5.15) into (3.11).

Imagine for a moment that we leave our original problem behind and only consider the functions $Y$ and $\sigma$
that satisfy the two Painlev\'e equations (4.33) and (5.9) with the appropriate initial conditions. Then our
orthogonal polynomials $P_n(z,t)$ satisfies the linear second order ode
\bea
\Psi''(z)+\textsf{P}(z)\Psi'(z)+\textsf{Q}(z)\Psi(z)=0,
\eea
where
\bea
\textsf{P}(z)&:=&\frac{1+\al}{z-1}+\frac{1+\bt}{z-1}-t-\frac{1}{z-[1+Y(2t)]/[1-Y(2t)]}\\
\textsf{Q}(z)&:=&-\frac{1}{2(z-1)}\left[\sigma(2t)+n(\al+1)-\frac{1-Y(2t)}{2}\frac{d}{dt}\sigma(2t)\right]\non\\
&+&\frac{1}{2(z+1)}\left[\sigma(2t)+n(\al+1+2t)-\left[\frac{1}{2}\frac{d}{dt}\sigma(2t)+n\right](1-Y(2t))\right]\non\\
&-&\frac{n+\frac{d}{dt}\sigma(2t)}{2}[1-Y(2t)]\left[\frac{1}{z-(1+Y(2t))/(1-Y(2t))}\right].
\eea
This is a deformation of the classical ode satisfied by the Jacobi polynomials. When $t=0$ this ode
reduces to a hypergeometric equation.

\setcounter{equation}{0}
\section{Toeplitz and Hankel determinants.}
In this section we introduce certain matrices that are combinations of finite Toeplitz and Hankel matrices.
There are identities that link these matrices directly to the Hankel moment matrices that appear in the first
section of this paper and define our  quantity $D_{n}(t).$ We will use these identities in some special
cases to get exact formulas for $D_{n}(t)$ and, as a by-product, find Painlev\'e type results for some other interesting
determinants. We include the Toeplitz/Hankel computations because as far as we know they are not written down explicitly in this form in any other place. 
However, the Case 2 was established already by two of the authors in  \cite[Sec.~2]{BE-LAA}.
The current derivation follows that in \cite{BE-LAA}.

Given a sequence
$\{a_k\}_{k=-\infty}^\infty$
of complex numbers, we associate the formal Fourier series
\bea\label{eqn6.1}
a(e^{i\theta })&=&\sum_{k=-\infty}^\infty a_k e^{i k \theta},\qquad e^{i\theta}\in \mathbb T.
\eea
The $n\times n$ Toeplitz and Hankel matrices
with the (Fourier) symbol $a$ are defined by
\bea\label{eqn6.2}
T_n(a) \;\;=\;\; \left(a_{j-k}\right)_{j,k=0}^{n-1},
\qquad
H_n(a) \;\;=\;\; \left(a_{j+k+1}\right)_{j,k=0}^{n-1}.\label{f.THn}
\eea
Usually $a$ represents an $L^1$-function defined on the unit circle $\T=\{z\in\C\,:\,|z|=1\}$,
in which case the numbers $a_k$ are the Fourier coefficients,
\bea\label{eqn6.3}
a_k &=& \frac{1}{2\pi}\int_{-\pi}^{\pi}a(e^{i\theta})
e^{-ik\theta}\,d\theta,\qquad k\in\Z.
\eea
Notice that while the matrices $H_{n}(a)$ are classically referred to as Hankel
matrices they are not the same as the Hankel moment matrices considered in the previous
sections of this paper.
To make the connection to Hankel matrices defined by moments we write
\begin{equation}\label{eqn6.3b}
H_{n}[b]  = (b_{j+k})_{j,k=0}^{n-1},\qquad b_k=\frac{1}{\pi} \int_{-1}^{1} b(x) (2x)^{j+k} dx,
\end{equation}
where $b(x)$ be an $L^1$-function defined on $[-1,1]$. Notice the difference in notation in comparison to (\ref{f.mu-i}) and (\ref{f.Dn-t}).
Our goal in this section is to prove four identities. Let $z = e^{i\theta}$. Then for each $n\ge1$ the following statements are true.
\begin{enumerate}
\item  If $
   a(e^{i\theta})= b(\cos\theta)(2+2\cos\theta)^{-1/2}(2-2\cos\theta)^{-1/2},$ then
   \[\,\,\, \det \Big(T_n(a)-H_n(z^{-1} a)\Big) =  \det H_n[b] .\]
\item If $
a(e^{i\theta})= b(\cos\theta)(2+2\cos\theta)^{-1/2}(2-2\cos\theta)^{1/2} ,$ then
\[\det \Big(T_n(a)+H_n(a)\Big) = \det H_n[b]. \]
\item If $
a(e^{i\theta})= b(\cos\theta)(2+2\cos\theta)^{1/2}(2-2\cos\theta)^{-1/2}$ then
\[\det \Big(T_n(a)-H_n(a)\Big) = \det H_n[b].\]
\item If $ a(e^{i\theta})= b(\cos\theta)(2+2\cos\theta)^{1/2}(2-2\cos\theta)^{1/2}$ then
\[\frac{1}{4}\det \Big(T_n(a)+H_n(z a)\Big) =  \det H_n[b].\]
 \vspace{.3cm}
 \end{enumerate}
In these identities the function $a$ is always even, which means in terms of its Fourier coeffcients that $a_k=a_{-k}$. Moreover, in these formulas we assume that $a\in L^1(\T)$, which implies that (and in Case 4 is equivalent to) $b\in L^1[-1,1]$. 
We also remark that Cases 2 and  3 can be derived from each other by making the substitutions
$a(e^{i\theta})\mapsto a(e^{i(\theta+\pi)})$ and $b(x)\mapsto b(-x)$.

Our interest in the above formulas stems from the circumstance that they allow us to use existing results 
\cite{BE-oam} 
on the asymptotics of the Toeplitz+Hankel determinants with well-behaved symbols $a$
in order to derive the asymptotics of the Hankel moment determinants.

In the above identities four types of finite symmetric Toeplitz+Hankel matrices as well as a finite Hankel moment matrix occur. These finite matrices can be obtained  from their (on-sided) infinite
matrix versions by taking the finite sections. It turns out that these infinite matrices are related to each other in a very simply way, namely they can be transformed into one another by multiplying with appropriate upper and lower triangular
(infinite) matrices form the left and right. These identities for the infinite matrices will be established
in the next theorem (and the remarks afterwards) in most general setting, where we do not assume that the symbols are $L^1$-functions.

Let us introduce the infinite matrices
$$
D_+=\left(\begin{array}{cccc} 
1\\-1&1\\&-1&1\\&&\ddots&\ddots\\
\end{array}\right),
\qquad
D_-=\left(\begin{array}{cccc} 
1\\ 1&1\\& 1&1\\&&\ddots&\ddots\\
\end{array}\right).
$$
These are just the well-known Toeplitz operators $D_\pm=T(1\mp z)$ and their transposes are denoted
by $D_\pm^T$. We also need the infinite diagonal matrix $R=\mathrm{diag}(\frac{1}{2},1,1,\dots)$.

\begin{theorem}\label{thm5}
For sequences of numbers $\{a_n\}_{n=-\iy}^\iy$, $\{a_n^+\}_{n=-\iy}^\iy$, $\{a_n^-\}_{n=-\iy}^\iy$, and $\{a_n^\#\}_{n=-\iy}^\iy$
satisfying
$$
a_n=a_{-n}, \quad a_n^+=a_{-n}^+, \quad a_n^-=a_{-n}^-, \quad a_n^\#=a_{-n}^\#, 
$$
define
\begin{equation}\label{eqn6.4}
\begin{array}{lcrclcr}
A &=& (a_{j-k}-a_{j+k+2})_{j,k=0}^\iy &\qquad&
A^+ &=& ( a_{j-k}^++a_{j+k+1}^+)_{j,k= 0}^\iy\\[2ex]
A^-  &=& ( a_{j-k}^--a_{j+k+1}^-)_{j,k= 0}^\iy &&
A^{\#}  &=& ( a_{j-k}^{\#}+a_{j+k}^{\#})_{j,k= 0}^\iy.
\end{array}
\end{equation}
Then the following holds true:
\begin{enumerate}
\item[(1)]
If $a^+_k=2a_k-a_{k-1}-a_{k+1}$, then $D_+A D_+^T=A^+$.
\item[(2)]
If $a^-_k=2a_k+a_{k-1}+a_{k+1}$, then $D_-A D_-^T=A^-$.
\item[(3)]
If $a^\#_k=2a^+_k+a^+_{k-1}+a^+_{k+1}$, then $D_- A^+D_-^T=RA^\#R.$
\item[(4)]
If $a^\#_k=2a^-_k-a^-_{k-1}-a^-_{k+1}$, then $D_+A^-D_+^T=RA^\#R.$
\end{enumerate}
Moreover, if we define a sequence $\{b_n\}_{n=0}^\iy$ and an infinite Hankel matrix by
\bq\label{f.b1}
b_n=\frac{1}{2}\sum_{k=0}^n a_{n-2k}^\# {n \choose k}, \qquad B=(b_{j+k})_{j,k=0}^\iy,
\eq
then 
\bq\label{eqn.6.8}
B=S_\# RA^\# R S_\#^T\quad\mbox{with}\quad S_\#=\left(
\begin{array}{cccccc}
{0\ch 0} \\[1ex]
0 & {1\ch 0}\\[1ex]
{2\ch1} & 0& {2\ch0} \\[1ex]
0 & {3\ch1} & 0 & {3\ch0} \\[1ex]
{4\ch2} & 0 & {4\ch1} & 0 & {4\ch0} \\[1ex]
\vdots &&&&&\ddots
\end{array}
\right).
\eq
\end{theorem}
{\em Proof:}
Before we start with the actual proof, we remark that the various products of the infinite matrices make sense in terms of the usual matrix
multiplication because the left and right factors are always (infinite) band matrices.

In order to prove the first statement (1) we consider the $(j,k)$-entries of the following (products of) infinite matrices and compute as follows:
\bq
\left[ D_+ A D_+^T\right]_{j,k} &=&
\left\{\ba{ll}
(2a_{j-k}-a_{j-k-1}-a_{j-k+1})-(a_{j+k+2}-2a_{j+k+1}+a_{j+k}) & \mbox{ if } j,k\ge1\\[1ex]
(a_{-k}-a_{-k+1})-(a_{k+2}-a_{k+1})& \mbox{ if } j=0,k\ge1\\[1ex]
(a_{j}-a_{j-1})-(a_{j+2}-a_{j+1}) & \mbox{ if } j\ge1, k=0\\[1ex]
a_0-a_2 & \mbox{ if } j=k=0
\ea\right.\nn\\[1ex]
&=&
\left\{\ba{ll}
a_{j-k}^+ + a^+_{j+k+1} & \mbox{ if } j,k\ge1\\[1ex]
a_{-k}^++a_{k+1}^+& \mbox{ if } j=0,k\ge1\\[1ex]
a_{j}^++a_{j+1}^+ & \mbox{ if } j\ge1, k=0\\[1ex]
a_0^+ +a_1^+ & \mbox{ if } j=k=0
\ea\right.
\quad=\quad \left[A^+\right]_{j,k} \nn
\eq
Herein we use the fact that $a_k-a_{k-1}-a_{k+2}+a_{k+1}=(2a_k-a_{k-1}-a_{k+1})+(2 a_{k+1}-a_{k}-a_{k+2})$,
a similar identity statement for $j$, and $a_0-a_2=(2 a_0 -2 a_1)+(2 a_1-a_0-a_2)$. Moreover, we use the assumption that 
$a_n=a_{-n}$.

Similarly, we compute the $(j,k)$-entry for the product appearing in (2):
\bq
\left[ D_- A D_-^T\right]_{j,k} &=&
\left\{\ba{ll}
(2a_{j-k}+a_{j-k-1}+a_{j-k+1})-(a_{j+k+2}+2a_{j+k+1}+a_{j+k}) & \mbox{ if } j,k\ge1\\[1ex]
(a_{-k}+a_{-k+1})-(a_{k+2}+a_{k+1})& \mbox{ if } j=0,k\ge1\\[1ex]
(a_{j}+a_{j-1})-(a_{j+2}+a_{j+1}) & \mbox{ if } j\ge1, k=0\\[1ex]
a_0-a_2 & \mbox{ if } j=k=0
\ea\right.\nn\\[1ex]
&=&
\left\{\ba{ll}
a_{j-k}^- - a^-_{j+k+1} & \mbox{ if } j,k\ge1\\[1ex]
a_{-k}^--a_{k+1}^-& \mbox{ if } j=0,k\ge1\\[1ex]
a_{j}^--a_{j+1}^- & \mbox{ if } j\ge1, k=0\\[1ex]
a_0^--a_1^- & \mbox{ if } j=k=0
\ea\right.
\quad=\quad \left[A^-\right]_{j,k}.\nn
\eq
Here we used $a_k+a_{k-1}-a_{k+2}-a_{k+1}=(2a_k+a_{k-1}+a_{k+1})-(2 a_{k+1}+a_{k}+a_{k+2})$,
a similar identity for $j$, and $a_0-a_2=(2 a_0 +2 a_1)-(2 a_1+a_0+a_2)$, and again $a_n=a_{-n}$.

As for statement (3) we consider
\bq
\left[ D_- A^+ D_-^T\right]_{j,k} &=&
\left\{\ba{ll}
(2a_{j-k}^++a_{j-k-1}^++a_{j-k+1}^+)+(a_{j+k+1}^++2a_{j+k}^++a_{j+k-1}^+) & \mbox{ if } j,k\ge1\\[1ex]
(a_{-k}^++a_{-k+1}^+)+(a_{k+1}^++a_{k}^+)& \mbox{ if } j=0,k\ge1\\[1ex]
(a_{j}^++a_{j-1}^+)+(a_{j+1}^++a_{j}^+) & \mbox{ if } j\ge1, k=0\\[1ex]
a_0^++a_1^+ & \mbox{ if } j=k=0
\ea\right.\nn\\[1ex]
&=&
\left\{\ba{ll}
a_{j-k}^{\#}+  a^{\#}_{j+k} & \mbox{ if } j,k\ge1\\[1ex]
\frac{1}{2}(a_{-k}^{\#}+a_{k}^{\#})& \mbox{ if } j=0,k\ge1\\[1ex]
\frac{1}{2}(a_{j}^{\#}+a_{j}^{\#}) & \mbox{ if } j\ge1, k=0\\[1ex]
\frac{1}{4}(a_0^{\#}+a_0^{\#}) & \mbox{ if } j=k=0
\ea\right.
\quad=\quad \left[R A^{\#}R \right]_{j,k}.\nn
\eq
Again, we used that $a^+_{n}=a^+_{-n}$ and $a^\#_{n}=a^\#_{-n}$.

Statement (4) can be proven in the same way as statement (3). In fact, if we assume all the hypotheses in (1)--(4), then (4) follows
with a little algebra from the previous three statements (and the from fact that $D_+$ and $D_-$ commute).

In order to prove formula (\ref{eqn.6.8}) first observe that 
$$
S_\#=(\xi(i,j))_{i,j=0}^\iy,\qquad \xi(i,j)=\left\{\begin{array}{cl}{i\ch \frac{i-j}{2}} &\mbox{ if } i\ge j \mbox{ and } i-j \mbox{ even}\\
0 & \mbox{otherwise.}\end{array}\right.
$$
Put $r_0=1/2$ and $r_n=1$ for $n\ge1$. Then the identity $B=S_\# R A^\# R S_\#^T$
can be rephrased as
\bq\label{eqn.6.9}
\frac{1}{2}\sum_{m=0}^{i+l} a_{i+l-2m}^\# {i+l\ch m} = \sum_{j=0}^i\sum_{k=0}^l \xi(i,j)\xi(l,k)r_jr_k (a_{j-k}^\#+a_{j+k}^\#)
\eq
to hold true for each $i,l\ge0$. These identities are valid if for each integer $s\ge0$, the coefficients for $a_{s}^\#=a_{-s}^\#$ are the same on the left and right hand side. 

First assume $s>0$. In the right hand side, the coefficient for $a_{s}^\#=a_{-s}^\#$ is equal to the sum $N_1+N_2+N_3$, where
\bq
N_1 &=& \sum_{\scriptsize\begin{array}{c}{\scriptsize 0\le j\le i}\\ 0\le k\le l\\ s=j-k\end{array}}   \xi(i,j)\xi(l,k)r_jr_k 
=  \sum_{\scriptsize\begin{array}{c}{\scriptsize 0\le u\le i/2}\\ 0\le v\le l/2\\ s=i-2u-l+2v \end{array}}  {i\ch u}{l\ch v}r_{i-2u}\,r_{l-2v},\nn
\\
N_2 &=& \sum_{\scriptsize\begin{array}{c}{\scriptsize 0\le j\le i}\\ 0\le k\le l\\ s=k-j\end{array}}   \xi(i,j)\xi(l,k)r_jr_k 
=  \sum_{\scriptsize\begin{array}{c}{\scriptsize 0\le u\le i/2}\\ 0\le v\le l/2\\ s=-i+2u+l-2v \end{array}}  {i\ch u}{l\ch v}r_{i-2u}\,r_{l-2v},\nn
\\
N_3 &=& \sum_{\scriptsize\begin{array}{c}{\scriptsize 0\le j\le i}\\ 0\le k\le l\\ s=j+k\end{array}}   \xi(i,j)\xi(l,k)r_jr_k 
=  \sum_{\scriptsize\begin{array}{c}{\scriptsize 0\le u\le i/2}\\ 0\le v\le l/2\\ s=i-2u+l-2v \end{array}}  {i\ch u}{l\ch v}r_{i-2u}\,r_{l-2v}.\nn
\eq
Therein, we made a change of variables $j\mapsto u=(i-j)/2$ and $k\mapsto v=(l-k)/2$. The summation is over integer pairs $(u,v)$.
In the above expressions for $N_1$ and $N_2$ we make another change of variables
$v\mapsto l-v$ and $u\mapsto i-u$ to get the expressions
$$
\sum_{\scriptsize\begin{array}{c}{\scriptsize 0\le u\le i/2}\\ l/2\le v\le l\\ s=i-2u+l-2v \end{array}}  {i\ch u}{l\ch v}r_{i-2u}\,r_{2v-l}
\quad\mbox{ and }\quad
\sum_{\scriptsize\begin{array}{c}{\scriptsize i/2\le u\le i}\\ 0\le v\le l/2\\ s=i-2u+l-2v \end{array}}  {i\ch u}{l\ch v}r_{2u-i}\,r_{l-2v}.
$$ 
Since there are no indices $(u,v)$ satisfying $i/2\le u\le i$, $l/2\le v\le l$, and $s=i-2u+l-2v$, we obtain that $N_1+N_2+N_3$
equals
\bq\label{eqn.6.10}
\sum_{\scriptsize\begin{array}{c}{\scriptsize 0\le u\le i}\\ 0\le v\le l\\ s=i-2u+l-2v \end{array}}  {i\ch u}{l\ch v}=\xi(i+l,s)=\left\{
\begin{array}{cl} {i+l\ch \frac{i+l-s}{2}} & \mbox{ if } s\le i+l\mbox{ and } i+l-s \mbox{ even}\\
0 & \mbox{ otherwise.}\end{array}\right.\qquad
\eq
This is the desired result since the coefficient for $a_{s}^\#=a_{-s}^\#$ in the left hand side of (\ref{eqn.6.9})
is zero if $i+l-s$ is odd and $$\frac{1}{2}\left({i+l\ch \frac{i+l-s}{2}}+{i+l\ch \frac{i+l+s}{2}}\right)= {i+l\ch \frac{i+l-s}{2}}$$
otherwise.

In the case $s=0$ the coefficient for the term $a_0^\#$ on the right hand side of (\ref{eqn.6.9}) equals $N:=N_1+N_3=N_2+N_3$, while it equals $\frac{1}{2}\xi(i+l,0)$ on the right hand side. 
The manipulation of the expressions $N_k$ can be done in the same way, with the only 
difference that in the end there are indices which $(u,v)$ satisfying $i/2\le u\le i$, $l/2\le v\le l$, $i+l=2(u+v)$. This corresponds to a term
$N_4$, which happen to be equal to $N_3$. Thus $N=\frac{1}{2}(N_1+\dots + N_4)$ with $N_1+\dots +N_4$ equaling (\ref{eqn.6.10}).
This settles the case $s=0$.

Hence we have shown that the identity  (\ref{eqn.6.9}) holds, and this implies formula (\ref{eqn.6.8}).
\hfill$\Box$
\vspace{4ex}

In regard to the first part of the theorem we remark that the hypotheses in (1)--(4) are compatible to each other in the sense that 
the hypotheses in (1) and (3), as well as those in (2) and (4) imply that 
$$
a_k^\#=2a_k-a_{k-2}-a_{k+2}.
$$
Correspondingly, we have
$$
D A D^T=RA^\#R \quad \mbox{ with }\quad D=D_+D_-=D_-D_+.
$$

Elaborating on formula (\ref{eqn.6.8}) we remark that assuming the hypotheses in (1)--(4), one can express
the coefficients $b_n$ in terms of $a^\pm_k$ and $a_k$ as well. We record the corresponding results form 
completeness sake:
\bq\label{f.1}
&&b_n = \sum_{k=0}^{n} {n \ch k}(a_{n-2k}^++a^+_{2n+1-k}),\quad B=S_+A^+S_+^T,\quad S_+=S_\#D_-
\\
\label{f.2}
&&b_n = \sum_{k=0}^{n} {n \ch k}(a_{n-2k}^-- a^-_{2n+1-k}),\quad B=S_-A^-S_-^T,\quad S_-=S_\#D_+
\\
\label{f.3}
&& b_n =\sum_{k=0}^n {n \ch k} (a_{n-2k}-a_{2n+2-k}),\quad B=SAS^T,\qquad S=S_\#D
\eq
The matrices $S_\pm$ and $S$ evaluate as follows:

$$
S_+=
\left(\ba{cccccc}
{0\ch 0}\\[1ex]
{1\ch0} & {1\ch 0}\\[1ex]
{2\ch1}&{2\ch0}& {2\ch0}\\[1ex]
{3\ch1}&{3\ch1}&{3\ch0}&{3\ch0}\\[1ex]
{4\ch2}&{4\ch1}&{4\ch1}&{4\ch0}&{4\ch0}\\[1ex]
\vdots&&&&&\ddots
\ea\right),
\qquad\quad
S_-=
\left(\ba{cccccc}
{0\ch 0}\\[1ex]
-{1\ch0} & {1\ch 0}\\[1ex]
{2\ch1}&-{2\ch0}& {2\ch0}\\[1ex]
-{3\ch1}&{3\ch1}&-{3\ch0}&{3\ch0}\\[1ex]
{4\ch2}&-{4\ch1}&{4\ch1}&-{4\ch0}&{4\ch0}\\[1ex]
\vdots&&&&&\ddots
\ea\right),
$$
$$
S=
\left(\ba{cccccc}
{0\ch 0}\\[1ex]
0 & {1\ch 0}\\[1ex]
{2\ch1}-{2\ch0} & 0& {2\ch0}\\[1ex]
0&{3\ch1}-{3\ch0}&0&{3\ch0}\\[1ex]
{4\ch2}-{4\ch1}&0&{4\ch1}-{4\ch0}& 0&{4\ch0}\\[1ex]
\vdots&&&&&\ddots
\ea\right).
$$
Finally we remark that the recurrence relation (\ref{f.b1}) allows to express the coefficients $a_n^\#$ in terms of $b_n$,
$$
a_0^\#=2b_0,\qquad a_n^\#=a^\#_{-n}=\sum_{k=0}^{[\frac{n}{2}]} (-1)^k b_{n-2k}\left(
{n-k\ch k} +{n-k-1\ch k-1}\right),\qquad n\ge1.
$$

Let us now proceed with establishing the identities for the determinants of the finite matrices.
We restrict to the cases where the symbols are $L^1$-functions because this is what is of interest to us.

\begin{theorem}\label{thm6}
Let $a, a^+,a^-,a^{\#}\in L^1(\T)$ be even, and let $b\in L^1[-1,1]$. Assume that 
\begin{enumerate}
\item
$a(e^{i\theta})= b(\cos\theta)(2+2\cos\theta)^{-1/2}(2-2\cos\theta)^{-1/2}$,
\item
$a^+(e^{i\theta})= b(\cos\theta)(2+2\cos\theta)^{-1/2}(2-2\cos\theta)^{1/2}$,
\item
$a^-(e^{i\theta})= b(\cos\theta)(2+2\cos\theta)^{1/2}(2-2\cos\theta)^{-1/2} $,
\item
$a^\#(e^{i\theta})= b(\cos\theta)(2+2\cos\theta)^{1/2}(2-2\cos\theta)^{1/2} $.
\end{enumerate}
Then, for each $n\ge1$,
\begin{equation}\label{eqn.6.4}
\begin{array}{l}
\det H_n[b]=
\det\Big( T_n(a)-H_n(z\iv a)\Big)=
\det\Big(T_n(a^{+})+H_n(a^+)\Big)\\[2ex]
\qquad=
\det\Big(T_n(a^{-})-H_n(a^-)\Big)=
\displaystyle \frac{1}{4}
\det\Big(T_n(a^{\#})+H(za^{\#})\Big).\end{array}
\end{equation}
\end{theorem}
{\em Proof.}
We first notice that the hypotheses on the coefficients stated in (1)--(4) of Theorem \ref{thm5}
can be rephrased in terms of the corresponding generating functions (see (\ref{eqn6.1}) and (\ref{eqn6.3})) as follows:
\begin{equation}\label{eqn6.13}
\begin{array}{lcrclcr}
a^+(z) &=&a(z)(1-z)(1-z\iv), &\qquad& a^-(z) &=&a(z)(1+z)(1+z\iv), \\[1ex]
a^\#(z) &=& a^+(z)(1+z)(1+z\iv), && a^\#(z) &=& a^-(z)(1-z)(1-z\iv).
\end{array}
\end{equation}
Here $z=e^{i\theta}\in\T$.
Incidentally, the relations between $a$, $a^+$, $a^-$, and $a^\#$ implied by the assumption 1.--4.\ above are precisely those in (\ref{eqn6.13}).

Now assume 4.\ and compute 
\bea
b_n &=& \frac{1}{\pi}\int_1^{-1} b(x)(2x)^{n}\, dx= \frac{1}{\pi}\int_0^{\pi} b(\cos \theta)(2\cos\theta)^n\sin(\theta)\, d\theta\nn \\
&=& \frac{1}{4\pi}\int_0^{2\pi} a^\#(e^{i\theta})(e^{i\theta}+e^{-i\theta})^n\, d\theta
=
\frac{1}{2}\sum_{k=0}^n a^\#_{n-2k}{n \ch k},\nn
\eea
which is precisely the condition (\ref{f.b1}).

In order to use the results of Theorem \ref{thm5} we take the finite sections of the various identities
(i.e., we consider the $n\times n$ upper-left corners of the infinite matrices),
$$
D_+ A D_+^T= A^+,\qquad D_-A D_-^T=A^-,
$$
$$
D_-A^+D_-^T=D_+A^-D_+^T=RA^\#R,\qquad B=S_\#RA^\#RS_\#^T,
$$
and then take the determinants. The crucial point is that $D_+$, $D_-$, and $S_\#$ are
lower-trinagular and have ones on their diagonals. The diagonal matrices $R$ give a factor
$\frac{1}{4}$ in the determinants. Now it just remains to check that the finite sections of the infinite matrices (\ref{eqn6.4}) and (\ref{f.b1}) are indeed the matrices occuring in (\ref{eqn.6.4}).
But this follows from the definitions (\ref{eqn6.2}) and (\ref{eqn6.3b}).
\hfill$\Box$
\vspace{3ex}

It is apparent from the proof that if we are only interested in an identity between two types of determinants featuring (\ref{eqn.6.4}), then it is enough to assume that only the corresponding
symbols are $L^1$-functions and that the appropriate relationships between these symbols
hold (see also (\ref{eqn6.13})). For instance, if we assume $a^+,a^-\in L^1(\T)$ and
$$
a^+(z)(1+z)(1+z\iv)=a^-(z)(1-z)(1-z\iv)
$$
then we can conclude that 
\begin{equation}
\det\Big(T_n(a^{+})+H_n(a^+)\Big)
=\det\Big(T_n(a^{-})-H_n(a^-)\Big).
\end{equation}
By the way, this relationship between these two type of determinants is not the trivial one featuring the ``equivalence'' between the Cases 2 and 3, which has been pointed out earlier.

\setcounter{equation}{0}
\section{Results from the Toeplitz theory}
The idea for this section is that if the $\alpha$ and $\beta$ are any combination of
$\pm\frac{1}{2},$ then we may choose an operator of the form $T_{n}(a) + H_{n}(b)$ from our
list of identities 1.--4.\ that has, in a certain sense, a nice symbol  and find an explicit formula
for  the determinants of the associated Hankel matrices, $H_n[b]$. This is because for these values
of the parameters and the right choice of operator, we lose the square root singularities.
Fortunately in \cite{BE-oam} exact formulas for the types of Toeplitz plus
Hankel determinants that appear in the previous theorem were found. If we specialize the results to
the cases at hand we can state the exact formula of the determinants of the matrices $H_n[b]$. The four
different determinants all have the form:
\begin{equation}\label{eqn7.16}
 G[a]^{n}F[a] \det (I + Q_{n} K Q_{n} ),
\end{equation}
where $F[a]$ and $G[a]$ are certain constants that depend on our choice of parameters for $\alpha$ and $\beta.$
The last operator determinant involves orthogonal projections $Q_n=I-P_n$, where 
the projections $P_n$ acting on $\ell^2(\mathbb{Z}_+)$, $\mathbb{Z}_+=\{0,1,\dots,\}$, are defined by
$$
P_{n}(a_{0}, a_{1}, \dots ) = (a_{0}, a_{1}, \dots, a_{n-1}, 0, 0,\dots).$$
The operator $K$, acting on $\ell^2(\mathbb{Z})$, is a certain (trace class) semi-infinite Hankel operator.

The precise reference for the result (\ref{eqn7.16}) is Proposition 4.1 and  the remark afterwards in \cite{BE-oam}. Propositions 3.1 and 3.3 in \cite{BE-oam} also have to be consulted. For sake of clarification we remark that our
cases 1.-4. correspond to the cases I-IV in \cite{BE-oam} as follows:
1.=III, 2.=I, 3.=II, 4.=IV, where in Case 4, the operators differ by a constant.

In our case the symbol is (up to a constant) $a(e^{i\theta})=e^{-t\cos\theta}$ whence
$\psi=a_+\iv(e^{i\theta})\tilde{a}_+(e^{i\theta})=e^{i t \sin\theta}$, which occurs in the definition of the operator $K$. The Fourier coefficients $\psi_k$ ($k\ge0$) are precisely equal to the value of 
the Bessel function $J_{k}(t)$ of order $j$ with the argument $t$. The precise description of $K$ is as follows:

\begin{enumerate}
\item Let $\alpha = \frac{1}{2}, \beta = \frac{1}{2},$ then
$ K = K_{1}$ where $K_{1}$ has $(j,k)$-entry $-J_{j+k+2}(t)$.
\item Let $\alpha = -\frac{1}{2}, \beta = \frac{1}{2},$ then
$ K = K_{2}$ where $K_{2}$ has $(j,k)$-entry $J_{j+k+1}(t)$.
\item Let $\alpha = \frac{1}{2}, \beta = -\frac{1}{2},$ then
$ K = K_{3}$ where $K_{3}$ has $(j,k)$-entry $-J_{j+k+1}(t)$.
\item Let $\alpha = -\frac{1}{2}, \beta = -\frac{1}{2},$ then
$ K = K_{4}$ where $K_{4}$ has $(j,k)$-entry $J_{j+k}(t)$.
\end{enumerate}
Here $j , k \geq 0.$   It is known that the operator $K$ is trace class.
This is not hard to see since for fixed $t$ the entries in the Hankel matrix tend to zero very rapidly.
We state the four cases below. In all cases our function $a$ in the previous identities is $e^{-t\cos \theta}$ times
 a factor of a power of $2$.

\begin{theorem} Let $b(x) = (1-x)^{\alpha}(1+x)^{\beta} e^{-tx}$.
\begin{enumerate}
\item Let $\alpha = \frac{1}{2}, \beta = \frac{1}{2},$ then
\[ \det H_n[ b] = 2^{-n}e^{t^{2}/8} \det\left(I + Q_{n} K_{1} Q_{n}\right).\]
\item Let $\alpha = -\frac{1}{2}, \beta = \frac{1}{2},$ then
\[ \det H_n[b] = e^{t^{2}/8-t/2} \det\left(I + Q_{n} K_{2} Q_{n}\right).\]
\item Let $\alpha = \frac{1}{2}, \beta = -\frac{1}{2},$ then
\[ \det H_n[ b] = e^{t^{2}/8+t/2} \det\left(I + Q_{n} K_{3} Q_{n}\right).\]
\item Let $\alpha = -\frac{1}{2}, \beta = -\frac{1}{2},$ then
\[ \det H_n[ b] = 2^{n-1}e^{t^{2}/8} \det\left(I + Q_{n} K_{4} Q_{n} \right).\]
\end{enumerate}
\end{theorem}
This does not quite give us the identity of the original $D_{n}(t)$ since the above Hankel was defined
with some extra constants of $\pi$ and $2$. So first we adjust for these to yield the following.
\begin{theorem} Let $b(x) = (1-x)^{\alpha}(1+x)^{\beta} e^{-tx}$.
\begin{enumerate}
\item Let $\alpha = \frac{1}{2}, \beta = \frac{1}{2},$ then
\[ D_{n}(t)  = 2^{-n(n+1)}(2\pi)^{n}e^{t^{2}/8} \det\left(I + Q_{n} K_{1} Q_{n} \right).\]
\item Let $\alpha = -\frac{1}{2}, \beta =  \frac{1}{2},$ then
\[ D_{n}(t) = 2^{-n^{2}}(2\pi)^{n}e^{t^{2}/8-t/2} \det\left(I + Q_{n} K _{2}Q_{n} \right).\]
\item Let $\alpha = \frac{1}{2}, \beta = -\frac{1}{2},$ then
\[ D_{n}(t) = 2^{-n^{2}}(2\pi)^{n}e^{t^{2}/8+t/2} \det\left(I + Q_{n} K_{3} Q_{n} \right).\]
\item Let $\alpha = -\frac{1}{2}, \beta = -\frac{1}{2},$ then
\[ D_{n}(t) = 2^{-n(n-1)-1}(2\pi)^{n}e^{t^{2}/8} \det\left(I + Q_{n} K_{4} Q_{n} \right).\]
\end{enumerate}
\end{theorem}
Since  $Q_{n}$ tends to zero strongly and the operator $K$ is trace class, the term
$\det\left(I + Q_{n} K Q_{n}\right)$ tends to one and the asymptotics are given by the previous factors
in each case of the above result.

More precisely, we obtain the following result.

\begin{theorem} Let $b(x) = (1-x)^{\alpha}(1+x)^{\beta} e^{-tx}$.
\begin{enumerate}
\item Let $\alpha = \frac{1}{2}, \beta = \frac{1}{2},$ then
\[ D_{n}(t)  \sim 2^{-n(n+1)}(2\pi)^{n}e^{t^{2}/8}. \]
\item Let $\alpha = -\frac{1}{2}, \beta = \frac{1}{2},$ then
\[ D_{n}(t) \sim  2^{-n^{2}}(2\pi)^{n}e^{t^{2}/8-t/2}. \]
\item Let $\alpha = \frac{1}{2}, \beta = -\frac{1}{2},$ then
\[ D_{n}(t) \sim  2^{-n^{2}}(2\pi)^{n}e^{t^{2}/8+t/2}. \]
\item Let $\alpha = -\frac{1}{2}, \beta = -\frac{1}{2},$ then
\[ D_{n}(t) \sim  2^{-n(n-1)-1}(2\pi)^{n}e^{t^{2}/8}. \]
\end{enumerate}
\end{theorem}
If we expand  $\det (I + Q_{n}K_{1}Q_{n})$ using the fact that $\log \det (I + A) = \mbox{tr} \log (I+A)$
using just the first couple of terms it seems reasonable to conjecture that, for example,
\[ D_{n}(t)  \sim 2^{-n(n+1)}(2\pi)^{n}\re^{t^{2}/8}\:
\re^{\frac{(t/2)^{2n+2}}{\Gamma(2n+3)} + {\rm O}\left(\frac{1}{\Gamma(2n+4)}\right)}. \]
Similar conjectures can be made in the other cases.

Before ending this section, we conjecture, with the aid the linear statistics formula in \cite{Chen+Lawrence} and \cite{Basor+Chen1}
obtained through the heuristic Coulomb Fluid approach \cite{Chen+Coul}, that for "general" values $\al$ and $\bt$ and for
large $n$
$$
\log\left(\frac{D_n(t)}{D_n(0)}\right)\sim
\frac{t^2}{8}+(\al-\bt)t,
$$
where
$$
D_n(0)\sim 2^{-n(n+\al+\bt)}\:n^{(\al^2+\bt^2)/2-1/4}\:(2\pi)^n\:
\frac{G(\frac{1+\al+\bt}{2})G^2(\frac{2+\al+\bt}{2})G(\frac{3+\al+\bt}{2})}
{G(1+\al+\bt)G(1+\al)G(1+\bt)}.
$$
Here $G(z)$ is the Barnes $G$-function \cite{bar}.

 Finally, we have as a consequence of the previous sections the
following remark: Let $\alpha = \frac{1}{2}, \beta = \frac{1}{2},$
and let
 $$\phi = t\frac{d}{dt} \log ( I   + Q_{n} K_{1}(t/2) Q_{n} ).$$
Then the function
$$\phi(t) = \frac{t^{2}}{16} + \frac{t}{2}\textsf{p}_{1}(n,\frac{t}{2})$$ and thus also
satisfies a related Painlev\'{e} equation. A similar expression can be obtained for
the other three cases. This once again demonstrates that the most fundamental quantity
in the theory is the coefficient $\textsf{p}_{1}(n,t).$

\end{document}